\documentclass[aps,prb,twocolumn,superscriptaddress,showpacs]{revtex4}
\usepackage{amsmath}    % need for subequations
\usepackage{graphicx}   % need for figures
\usepackage{verbatim}   % useful for program listings
\usepackage{color}      % use if color is used in text
\usepackage{subfigure}  % use for side-by-side figures
\usepackage[hypertex]{hyperref}   % use for hypertext links, including those to external documents and URLs

\usepackage{times}

\begin{document}

\title{Numerical study of Kondo impurity models with strong potential scattering:\\
-- reverse Kondo effect and antiresonance --
}

\author{Annam\'{a}ria Kiss}
\email{akiss@szfki.hu}
\affiliation{Budapest University of Technology and Economics, Institute
of Physics and Condensed Matter Research Group of the Hungarian Academy of
Sciences, H-1521 Budapest, Hungary}

\affiliation{Research Institute for Solid State Physics and Optics, P.O. Box 49, H-1525 Budapest, Hungary}

\author{Yoshio Kuramoto}
\affiliation{Department of Physics, Tohoku University, Sendai, 980-8578, Japan}

\author{Shintaro Hoshino}
\affiliation{Department of Physics, Tohoku University, Sendai, 980-8578, Japan}

\begin{abstract}
Accurate numerical results are derived for transport properties of Kondo impurity systems with potential scattering and orbital degeneracy. Using the continuous-time quantum Monte Carlo (CT-QMC) method, static and dynamic physical quantities are derived in a wide 
temperature
range across the Kondo temperature $T_{\rm K}$.
With 
strong
potential scattering, the resistivity tends to decrease with decreasing temperature, in contrast to the ordinary Kondo effect. 
Correspondingly, the quasi-particle density of states obtains the antiresonance around the Fermi level.
Thermopower also shows characteristic deviation from the standard Kondo behavior, while magnetic susceptibility follows the universal temperature dependence even with 
strong
potential scattering.
It is found that 
the $t$-matrix in the presence of potential scattering
is not a relevant quantity for the Friedel sum rule, for which
a proper limit of the $f$-electron Green's function is introduced. 
The optical theorem is also discussed in the context of Kondo impurity models with potential scattering.
It is shown that optical theorem holds not only in the Fermi-liquid range but also for large energies,
and therefore
is less restrictive than the Friedel sum rule.
\end{abstract}

\maketitle

\section{Introduction}

A single magnetic impurity embedded into the sea of conduction electrons shows Kondo effect. 
Although this problem has been almost continuously studied during the last 40 years, dynamic and magnetic properties of the Kondo and related models still attract great interest in condensed matter physics. 
Much less attention is paid to anomalous {\it decrease} of resistivity with lowering temperature since such decrease can be caused by many different mechanisms. 
The first attention to this problem is traced back to the 60's, when resistivity of dilute Fe and Cu alloys in Rh matrix revealed a new type of anomaly at low temperatures \cite{coles64}.
Namely, it was observed that resistivity 
decreases with decreasing temperature in these compounds.
Obviously a ferromagnetic exchange interaction of a localized spin and conduction electrons is the first candidate. 
In fact, later study on dilute Gd and Nd impurities in some La alloys such as LaAl$_2$ \cite{lieke78}, and LaSn$_3$ \cite{schmid80} explained the resistivity anomaly in terms of the ferromagnetic exchange model \cite{lieke80}, and was referred to as the ``reverse Kondo effect".   

In dilute Rh alloys, however, the measured susceptibility indicates antiferromagnetic exchange interaction in these materials in contrast with the behavior of the resistivity.
As alternative interpretation, Fischer found that a strong potential scattering can change sign of the Kondo logarithmic term in the resistivity \cite{fischer67}.
The treatment has been much extended and deepened by Kondo \cite{kondo-1968}, who
uses the scattering phase shift $\delta_{v}$ of conduction electrons
at the Fermi surface.
The strength $v$ of the potential scattering is related to the phase shift
by $\tan \delta_{v} = -\pi v \rho_c$ where
$\rho_c$ is the density of states 
at the Fermi surface.
The leading logarithmic term in the electric resistivity changes sign when $|\delta_{v}|$ exceeds 
$\pi/4$, defining a critical value for the potential scattering as $v_{\rm cr} \equiv  
1/(\pi \rho_c)$. The range $|v|>v_{\rm cr}$ is called reverse Kondo range. In this range
the resistivity decreases with decreasing temperature as a consequence of the strong potential scattering.
Regarding thermodynamic properties,  Kondo showed that 
the effect of ordinary scattering is entirely absorbed into an effective exchange interaction 
$\widetilde{J}=J \cos^2 \delta_{v}$. 
Correspondingly, 
the low-temperature energy scale is characterized by temperature $T_{\rm K} = D {\rm e}^{1/(2\rho_c \widetilde{J})}$ associated with the effective exchange interaction $\widetilde{J}$.

Since ordinary scattering events are always present in real systems,
the Kondo problem with strong potential scattering might have relevance also in other
compounds that show Kondo effect.
For example, the question of relevance of ordinary scattering in URu$_{2}$Si$_{2}$ arises, because:
(i) recent STM experiments have found that the density of electronic states shows Fano lineshape, 
i.e. antiresonance, in the 
normal phase,\cite{schmidt-2010} and
(ii) in the dilute system U$_{x}$Th$_{1-x}$Ru$_{2}$Si$_{2}$ the resistivity decreases with decreasing temperature.\cite{amitsuka-1994}
 
In this paper we study 
the effect of strong potential scattering on physical properties of the Kondo impurity.
In order to deal with the Kondo effect 
beyond the weak coupling regime,
the continuous-time quantum Monte Carlo (CT-QMC) method is employed
\cite{otsuki-2007, otsuki-2009a, otsuki-2009b}.
In the CT-QMC simulation it is most convenient to take
the $N$-component Coqblin-Schrieffer (CS) model\cite{coqblin-schrieffer-1969, otsuki-2007} with potential scattering.  The Hamiltonian is
given by 
\begin{eqnarray}
{\cal H}_{\rm CS} [v_{\rm CS}] &=& \sum_{{\bf k} m} \varepsilon_{\bf k} c_{{\bf k} m}^{\dag}c_{{\bf k} m}^{\phantom{\dag}}\nonumber\\
  &+& \sum_{m m^{\prime}}\left( J f^{\dag}_{m}f_{m^{\prime}}^{\phantom{\dag}} + v_{\rm CS}\delta_{m m^{\prime}} \right) c_{m^{\prime}}^{\dag}c_{m}^{\phantom{\dag}},\label{kondom}
\end{eqnarray}
where $c_{{\bf k}m}^{\dag}$ and $f_{m}^{\dag}$ are creation operators of conduction and localized electrons, respectively, at the impurity site with SU$(N)$ index $m = 1,\ldots, N$.
The constraint
$
\sum_m f_{m}^{\dag}f_{m}^{\phantom{\dag}} =1
$
is imposed, which removes the charge degrees of freedom.
The annihilation operator $c_{m}$ in the Wannier representation is related to 
$c_{{\bf k}m}$ by 
$c_{m}=N_{0}^{-1/2}\sum_{{\bf k}}c_{{\bf k}m}$ with $N_{0}$ being the number of lattice sites.  
We observe the relation
\begin{eqnarray}
\sum_{m m^{\prime}}  
f^{\dag}_{m}f_{m^{\prime}}^{\phantom{\dag}} 
c_{m^{\prime}}^{\dag}c_{m}^{\phantom{\dag}} &=& 
\sum_{m m^{\prime}} \tilde{X}_{m m^{\prime}} 
c_{m^{\prime}}^{\dag}c_{m}^{\phantom{\dag}}  
+\frac 1N n_c \nonumber \\ 
&& \mathop\Rightarrow_{N=2}
2{\bf S}_{f}\cdot {\bf s}_{c} + \frac 12 n_c 
,\label{eqkondo}
\end{eqnarray}
where 
$\tilde{X}_{m m^{\prime}} \equiv 
f^{\dag}_{m}f_{m^{\prime}}^{\phantom{\dag}} -\delta_{m m'} / N
$
are SU$(N)$ generators,
and
${\bf s}_{c}$ and $n_c$ are spin and charge density operators of conduction electrons at the impurity site.
The SU$(N)$ Kondo Hamiltonian ${\cal H}_{\rm K}[v]$
with potential scattering $v$ 
is introduced by the relation
\begin{equation}
{\cal H}_{\rm CS} [v_{\rm CS}]
= {\cal H}_{\rm K} [v=v_{\rm CS}+J/N],
\end{equation}
in view of eq.~(\ref{eqkondo}).
Some typical cases of the model given by Hamiltonian (\ref{kondom}) are
\begin{itemize}
\item[ (i) ] 
The conventional CS model with $v_{\rm CS}=0$, or $v=J/N$;
\item[ (ii) ] 
The SU$(N)$ Kondo model with $v = 0$, or $v_{\rm CS}=-J/N$; 
\item[ (iii) ] 
Reverse Kondo range with $|v| > v_{\rm cr} =1/(\pi \rho_c)$.
\end{itemize}

On the basis of accurate numerical results for strong potential scattering, 
we investigate the reverse Kondo range in detail.
Furthermore, properties are studied by changing the orbital degeneracy $N$ for the CS model.
This paper is organized as follows. In Section~\ref{section-susceptibility} numerical results for the magnetic susceptibility are presented.
The characteristics of the impurity $t$-matrix are discussed in Section~\ref{section-transport}.
Furthermore, numerical results are given for transport properties.
Section~\ref{section-fsr} is devoted to discussion of quasi-particle properties including the Friedel sum rule and optical theorem.
The summary of this paper will be given in Section~\ref{section-summary}.

\section{Magnetic susceptibility and universality}\label{section-susceptibility}

First, let us discuss the behavior of the magnetic susceptibility for a given value of the orbital degeneracy $N$. 
The static susceptibility is obtained from the imaginary time data by integration as
\begin{eqnarray}
\chi(T) = \int_{0}^{\beta} d\tau \chi(\tau) =  \int_{0}^{\beta} d\tau \langle T_{\tau} M^{\rm H}(\tau) M\rangle,
\end{eqnarray}
where the dipole moment $M$ is given by $M=\sum_{\alpha} m_{\alpha} f^{\dag}_{\alpha}f_{\alpha}^{\phantom{\dag}}$ with coefficients $m_{\alpha}$ chosen as $\sum_{\alpha} m_{\alpha} = 0$, and the superscript ${\rm H}$ denotes the Heisenberg picture\cite{otsuki-2007}. 
We use a constant density of states for the conduction electrons in the simulation as
\begin{eqnarray}
\rho_{c}(\varepsilon) = \rho_{0} \Theta (D - |\varepsilon|),
\end{eqnarray}
where $\rho_{0} = 1/(2D)$ with $D=1$ as a unit of energy.
In the numerical study, we determine the Kondo temperature from the low-temperature static susceptibility as
\begin{eqnarray}
T_{\rm K}^{-1} = \chi(T\rightarrow 0)/C_{N},
\end{eqnarray}
where $C_{N}$ is the Curie constant.
The critical strength $v_{\rm cr}$ is given by 
$v_{\rm cr} = 1/(\pi \rho_0) = 0.637$.

\begin{figure}
\centering
\includegraphics[totalheight=5.7cm,angle=0]{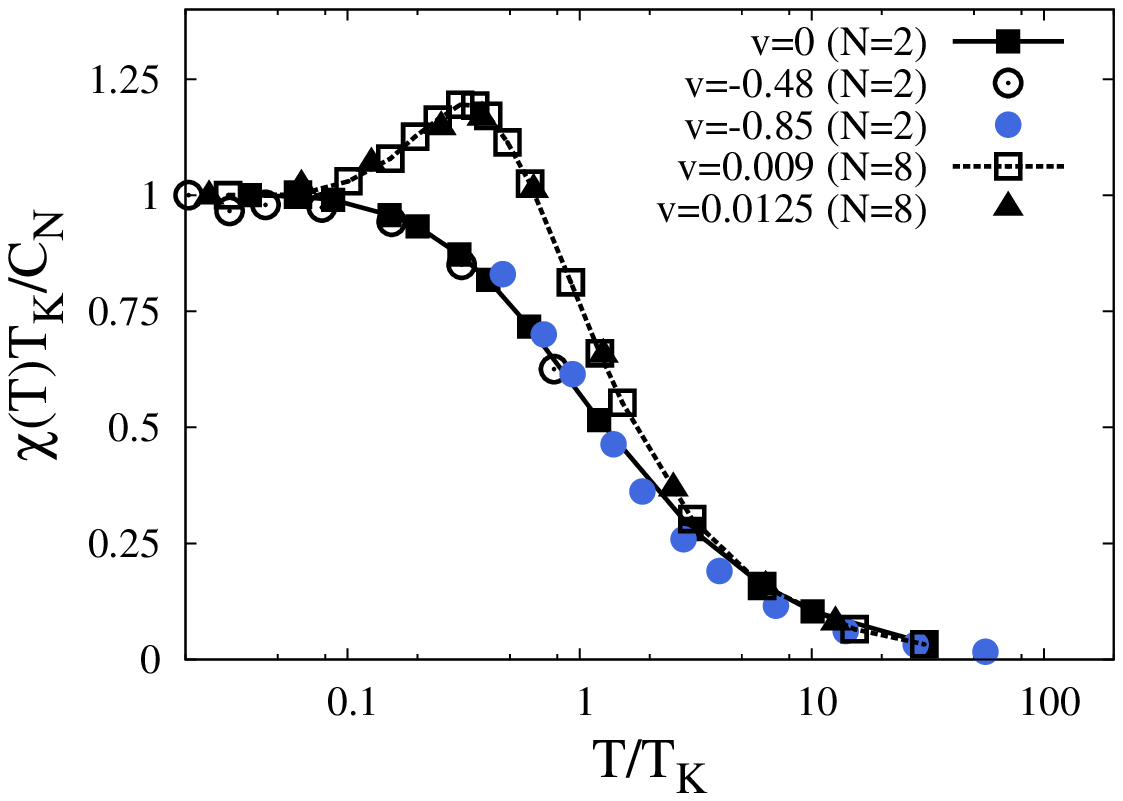}
\includegraphics[totalheight=5.7cm,angle=0]{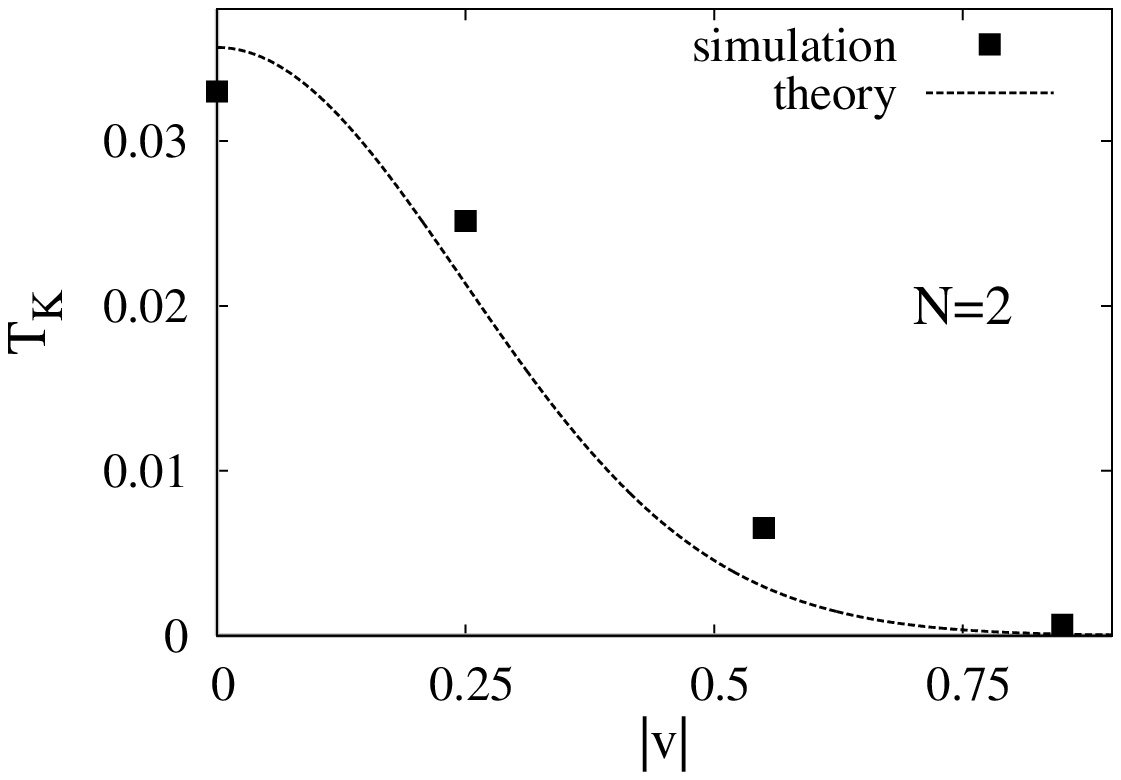}
\caption{{\sl Upper:} 
Temperature dependence of the static susceptibility for the 
SU$(N)$ Kondo model with potential scattering.
Potential scattering 
$v$ and exchange $J$ are chosen as 
$v=0, -0.85 \ (J=0.3)$, and $v= -0.48\ (J=0.44)$ for $N=2$, and 
$v= 0.009\ (J=0.075)$, and $v=0.0125\  (J=0.0125)$ for $N=8$.
{\sl Lower:} Kondo temperature for several values of potential scattering obtained in simulation. The theoretical result $T_{\rm K} = D {\rm e}^{1/(2\rho \widetilde{J})}$ is also shown as {\sl dashed line}. 
}
\label{susc}
\end{figure}

Figure \ref{susc} shows $\chi (T)$ of the SU$(N)$ Kondo (or CS) models 
with several values of the potential scattering $v$ and orbital degeneracy $N$
({\sl upper}), and $T_{\rm K}$ 
for different values of $v$ ({\sl lower}).
The result $T_{\rm K} = D {\rm e}^{1/(2\rho \widetilde{J})}$ obtained by Kondo \cite{kondo-1968} is also shown for comparison.
We observe in Fig.~\ref{susc} that the susceptibility shows universal behavior as a function of $T/T_{\rm K}$ independent of
the value of the potential scattering.
Note that the data with $v=-0.85$ for $N=2$ as shown by {\sl blue} symbols are in the reverse Kondo range with 
$\delta_{v}$ being larger than $\pi/4$. Even in this case, the temperature evolution of the magnetic susceptibility shows the universal behavior.

\begin{figure*}
\centering
\includegraphics[totalheight=4.8cm,angle=0]{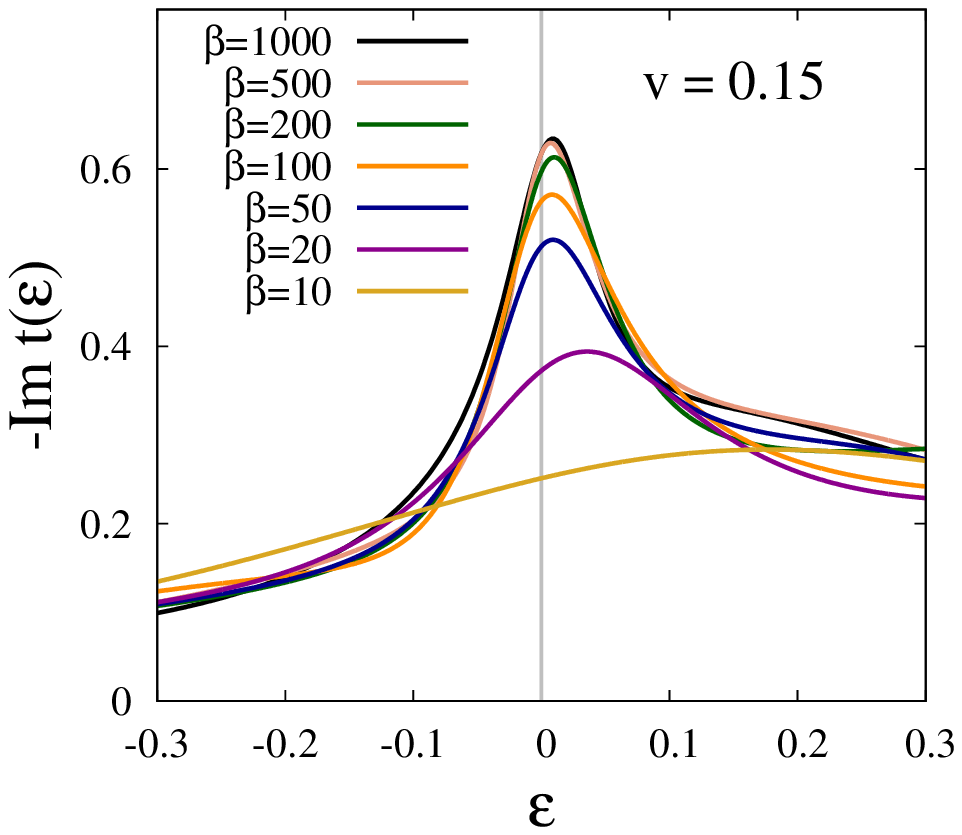}
\includegraphics[totalheight=4.8cm,angle=0]{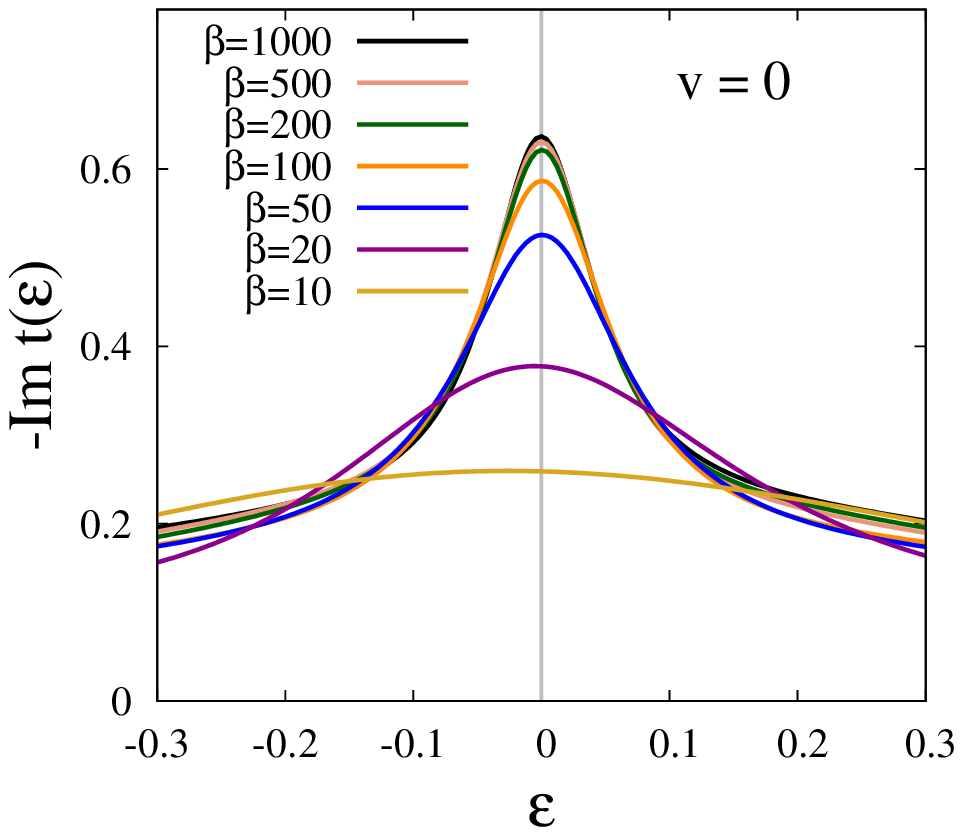}
\includegraphics[totalheight=4.8cm,angle=0]{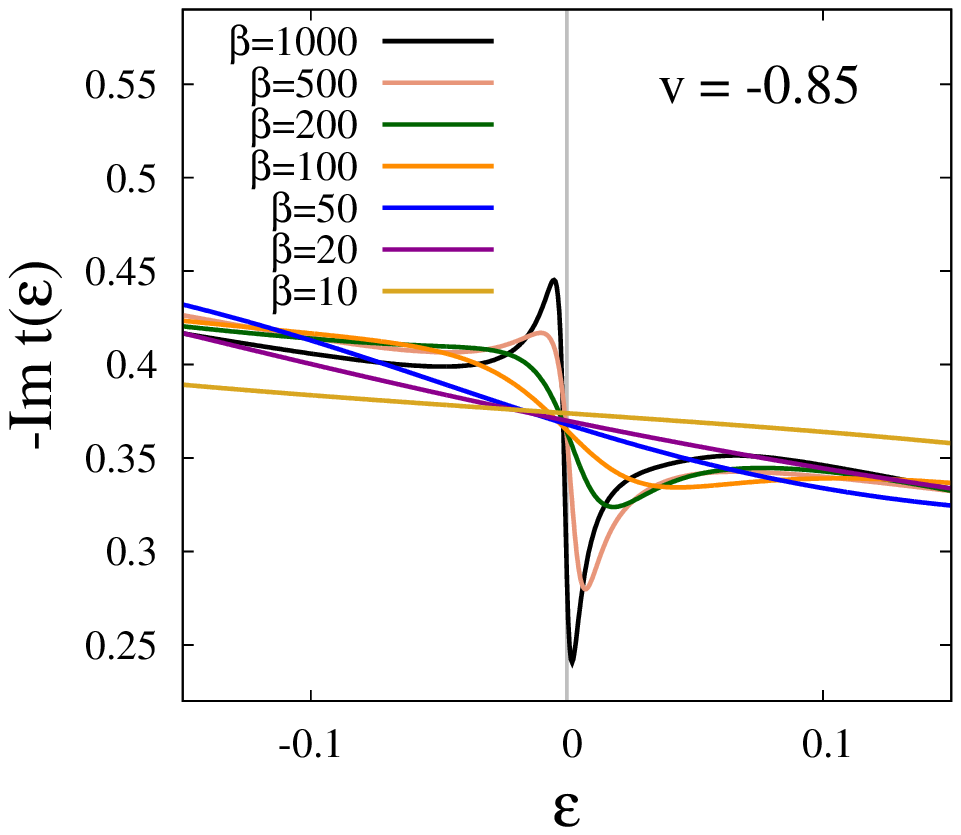}
\caption{
Energy dependence of -Im $t(\varepsilon)$
with potential scattering for 
$N=2$ and $J=0.3$. 
Potential scattering terms are chosen as $v=0.15$ ({\sl left}), $v=0$ ({\sl center}) and $v=-0.85$ ({\sl right}). Values $v=0.15$ and $v=0$ correspond to the CS and ordinary Kondo models, respectively.
}
\label{fdos}
\end{figure*}

Now, let us turn to discuss the properties by changing the orbital degeneracy $N$.
For $N=2$, 
the static susceptibility decreases monotonically with increasing temperature. On the other hand, for large values of the orbital degeneracy like $N=8$ in Fig.~\ref{susc}, the susceptibility first 
increases as the temperature is increased, 
and then decreases 
in accordance with the free moment behavior $\chi \sim 1/T$ for large temperatures.
The initial increase
can be understood in terms of the density of states
of quasi-particles.
To illustrate the mechanism of the increase,
let us consider the case of $v=0$ in 
the non-interacting Anderson model, which simulates qualitatively the 
quasi-particle density of states.
Using the Sommerfeld expansion 
\begin{eqnarray}
\chi(T)  &\sim & \int_{-\infty}^{\infty}  \rho_f(\varepsilon) \left( -\frac{\partial f(\varepsilon)}{\partial \varepsilon} \right) d\varepsilon \nonumber\\
 &=& 
\rho_f(0) \left[ 1+ \frac{\pi^2}{6} (k_{\rm B}T)^2 \frac{\rho_f^{\prime \prime}(0)}{\rho_f(0)} +{\cal O}(T^4)  \right]\label{somm-chi}
\end{eqnarray}
with
the $f$-electron density of states $\rho_f(\epsilon)$,
we obtain at low temperatures 
\begin{eqnarray}
\chi(T) \sim \rho_f (0) \left( 1+ \frac{\pi^2}{3} (k_{\rm B}T)^2  \frac{(3 \eta^2 - \Delta^2)}{(\eta^2 + \Delta^2)^2} +{\cal O}(T^4) \right),\label{somm-susc}
\end{eqnarray}
where $\eta$ and $\Delta$ are the shift and the width of the resonance peak appearing in the density of states $\rho_f(\varepsilon)$ at low temperatures.

For $N=2$ the resonance peak is centered at the Fermi energy, which gives $\eta = 0$. 
Therefore, the coefficient of $T^2$ in $\chi(T)$ is negative, i.e. the susceptibility decreases with increasing temperatures. 
Increasing the value of degeneracy $N$, the resonance moves to higher energy above the Fermi level, i.e. $\eta \sim T_{\rm K}$,
while its width narrows as $\Delta \sim T_{\rm K}/N$.\cite{hewson-book} Thus, the coefficient of $T^2$ in expression (\ref{somm-susc}) becomes positive, 
so that the susceptibility first shows increasing behavior as the temperature is increased.

\section{Transport coefficients}\label{section-transport}

\subsection{$t$-matrix and Fano lineshape}
We have already shown in eq.~(\ref{eqkondo}) that the Kondo and CS Hamiltonians are related to each other through a potential scattering term.
In the simulation for the CS model, instead of the bare Green's function $g$, another Green's function $g_{\rm CS}$ is used that absorbs
the potential scattering $v_{\rm CS}$: 
\begin{eqnarray}
g_{\rm CS} = g/(1-v_{\rm CS}g),
\end{eqnarray}
where
$g(z)=\sum_{\bf k}(z - \varepsilon_{\bf k} )^{-1}$.
Then the simulation gives the renormalized Green's function $G$ of conduction electrons.  We introduce a quantity
$t_{\rm CS}$ by the relation
\begin{eqnarray}
G = g_{\rm CS} + g_{\rm CS} t_{\rm CS} g_{\rm CS}.
\label{Gtcs}
\end{eqnarray}
On the other hand, the $t$-matrix $t$ of conduction electrons is defined by the relation
\begin{eqnarray}
G = g + g t g.\label{Gt}
\end{eqnarray}
By comparing eqs.~(\ref{Gtcs}) and (\ref{Gt}), we obtain $t$ from $t_{\rm CS}$ by the relation
\begin{eqnarray}
t = v_{\rm CS}/(1-v_{\rm CS}g) + t_{\rm CS}/(1-v_{\rm CS}g)^2.
\end{eqnarray}
In the special case of 
$v=v_{\rm CS}+J/2=0$, we recover eq.~(33) in Ref.~\onlinecite{otsuki-2007}.

For the moment, we concentrate on the case of $N=2$.
In the CT-QMC simulation, the $t$-matrix is derived in the imaginary-time domain. In order to obtain properties in the real energy
domain, analytic continuation of the numerical data is done by using Pad\'e approximation.
Figure~\ref{fdos} shows the energy
dependence of the impurity $t$-matrix for three different values of  the potential scattering at various temperatures.
To simplify the notation, we take the convention in this paper that 
energy including an infinitesimal imarginary part,  $\varepsilon + i\delta$,
is simply written as $\varepsilon$.
The {\sl left panel} of Fig.~\ref{fdos} with the value $v=0.15$ corresponds to the CS model, {\sl center panel} with $v=0$ to the ordinary single-channel Kondo model, while {\sl right panel} to a strong potential scattering $|v|>v_{\rm cr}=0.637$. 
In the case of the ordinary single-channel Kondo model the spectrum is symmetric with respect to the Fermi energy, because the model has particle-hole symmetry in this limit.
Increasing the value of the potential scattering, the Kondo peak first moves to higher frequencies above the Fermi energy. 
Finally, for strong potential scattering the spectrum becomes highly asymmetric showing an antiresonance around the Fermi level. 

Interpretation of the asymmetric spectrum with large $|v|$ can be provided 
in terms of the Anderson model, which reproduces the CS model (\ref{kondom}) in the limit of deep local electron level $\varepsilon_f$ and large Coulomb repulsion $U$ as compared with hybridization $V$.  
Namely we take the limits 
$\varepsilon_{f}\rightarrow -\infty$, 
$\varepsilon_{f}+U\rightarrow \infty$
and $V^2\rho_{0}\rightarrow \infty$,  
keeping the ratio $J=-2V^2\rho_{0}/\varepsilon_{f}$ finite. 
Now we construct the $f$-electron Green's function $G_{fv}$ of the Anderson model in the presence of potential scattering.
Let us first consider the pure case $v=0$.
Introducing the irreducible part $F(z)$,
we obtain the Green's function
\begin{eqnarray}
G_{f}(z) = F(z)[1+V^2g(z)G_{f}(z)].
\end{eqnarray}
In the presence of potential scattering, the $f$-electron Green's function $G_{fv}(z)$ satisfies the following relation
\begin{eqnarray}
G_{fv}(z) = F_{v}(z)[1+V^2g_{v}(z)G_{fv}(z)],
\end{eqnarray}
where $g_{v}(z) = g(z)/[1-v g(z)]$.

The $t$-matrix for the CS model is given by
\begin{eqnarray}
t(z) = \frac{v + V^2 F_{v}(z)}{1-g(z)[v + V^2 F_{v}(z)]} = t_{v}(z) + \frac{V^2 G_{fv}(z)}{[1-v g(z)]^2},\label{tz1}
\end{eqnarray}
where $t_{v}=v/(1-vg)$. It is clear from eq.~(\ref{tz1}) that the $t$-matrix reduces to
\begin{eqnarray}
t(z) \rightarrow V^2 G_{f}(z)
\end{eqnarray}
in the limit of $v=0$ as we expect. We derive from eq.~(\ref{tz1})
\begin{eqnarray}
V^2 G_{fv} = (1-vg)^2(t-t_{v})\label{tz2},
\end{eqnarray}
which is valid for any value of parameters.

As the simplest case, let us consider the non-interacting Anderson model with $U=0$.
Then we obtain 
$F(\varepsilon) = 1/(\varepsilon + i\delta - \varepsilon_f) \equiv 1/\xi$,
which is not affected by potentital scatteing.
The imaginary part of the $f$-electron Green's function is given by
\begin{align}
-{\rm Im} \,G_{fv}(\varepsilon) = \frac 1{V^2}\cdot
\frac{\pi \rho_{0} (v+V^2/\xi)^2 }{1 + \pi^2 \rho_{0}^2 (v+V^2/\xi)^2 }.
\label{eq5t}
\end{align}
Under the transformation $v \rightarrow -v$, the imaginary part of the Green's function given in eq.~(\ref{eq5t}) remains the same if
we put $
\varepsilon-\varepsilon_{f} \rightarrow 
\varepsilon_{f}-\varepsilon
$.
Physically it means that the antiresonance is reflected with respect to the energy $\varepsilon_{f}$ when $v$ changes sign.
Expression~(\ref{eq5t}) can be put into the standard form of the Fano lineshape \cite{fano-1961}. 
Introducing the dimensionless parameter $q$ 
by $1/q \equiv \pi v \rho_{0}$,
we rearrange the terms as follows:
\begin{align}
- \frac{V^2}{v}{\rm Im} \,  G_{fv} (\varepsilon) = 
\frac{1}{q+1/q}\cdot  \frac{(x+q)^2}{x^2+1},
\end{align}
where $x$ is the dimensionless energy defined by
\begin{eqnarray}
x  = \frac{v}{V^2}\left( q +\frac 1q \right) \xi+\frac 1q.
\end{eqnarray}
The degree of asymmetry is determined by the parameter $q$ that is independent of hybridization.

Expression~(\ref{eq5t}) describes the characteristics of the simulation results for $t(\varepsilon)$ shown in Fig.~\ref{fdos},
provided we put $\epsilon_f \sim 0$.
We note that $V^2G_{fv}(z)$ is not 
the same as the $t$-matrix $t(z)$ of the CS model as shown in eq.~(\ref{tz1}).
However, the characteristic lineshape comes almost from $V^2G_{fv}(z)$ since $t_v(z)$ and $1-v g(z)$ do not have strong dependence on $z$.
The asymmetric spectrum for strong potential scattering has interesting consequences with respect to transport properties such as the resistivity or thermopower. This problem is discussed in the next subsection.

\subsection{Relaxation time and transport coefficients}

We rely on the Boltzmann equation approach \cite{mahan-book} to derive  
transport coefficients.   Then the relaxation time $\tau(\varepsilon)$ is related to 
the $t$-matrix 
as\cite{hewson-book}
\begin{eqnarray}
\tau({\varepsilon})^{-1} = -2{\rm Im}\,t(\varepsilon).
\label{tau-t}
\end{eqnarray}
Let us introduce the integrals \cite{pruschke-2006}
\begin{eqnarray}
L_{n} =  \int_{-\infty}^{\infty} d\varepsilon \left(- \frac{\partial f(\varepsilon)}{\partial \varepsilon}  \right) \tau(\varepsilon)\varepsilon^{n},
\label{l-int}
\end{eqnarray}
in terms of which the conductivity $\sigma$, thermopower $S$ and thermal conductivity $\kappa$ are expressed as
\begin{eqnarray}
\sigma (T) &=& L_{0},\\
S(T)  &=& - \frac{1}{T} \frac{L_{1}}{L_{0}} ,\label{thermop}\\
\kappa (T)  &=& \frac{1}{T} \left(  L_{2} - \frac{L_{1}^{2}}{L_{0}} \right).
\end{eqnarray}
The integrals in eq.~(\ref{l-int}) are evaluated numerically with the CT-QMC data for the $t$-matrix.

\begin{figure}
\centering
\includegraphics[totalheight=6cm,angle=0]{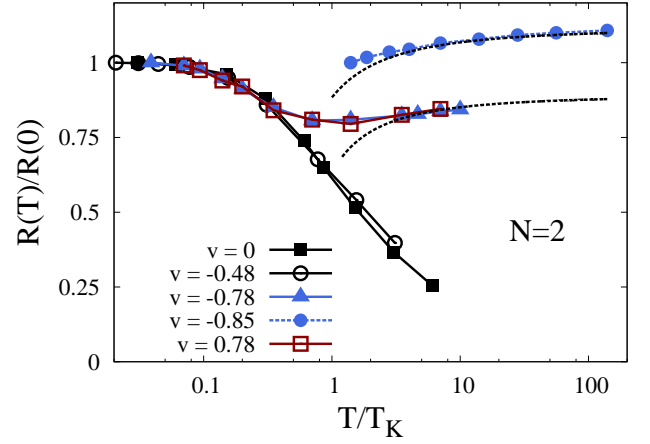}
\caption{Temperature dependence of the normalized resistivity for the Kondo model with potential scattering for $N=2$. 
Potential scattering $v$ and exchange $J$ are chosen as 
$v=0, -0.85 \ (J=0.3)$, and $v=0.78, -0.48, -0.78 \ (J=0.44)$.
 The {\sl dashed line} corresponds to Kondo's result for the resistivity given in Ref.~\onlinecite{kondo-1968}.}\label{res-tk}
\end{figure}

Although many theoretical attempts were made to derive the transport properties for the Kondo problem analytically in the whole temperature range, none of these attempts was successful. However, there are correct results in limiting cases such as Hamman's formula\cite{hamann-1967} for $T\gg T_{\rm K}$, Fermi-liquid results\cite{Nozieres-1974} for $T\ll T_{\rm K}$, or expressions for large values of the orbital degeneracy $N$.\cite{houghton-1987}
In the local Fermi-liquid range at low temperatures, the resistivity $R(T)$
shows the following temperature dependence:
\begin{align}
R(T)/R(0) = 1-\alpha \left( T/T_{\rm K} \right)^2, \label{res2}
\end{align}
where we have used the relation
$R(T)/R(0) = \sigma (0)/\sigma (T)$, and 
$\alpha$ is a numerical coefficient.
For large values of the orbital degeneracy $N$, the 1/$N$ expansion gives the coefficient $\alpha$ in eq.~(\ref{res2}) as \cite{houghton-1987}
\begin{eqnarray}
\alpha = \pi^2 \left( 1-\frac{8}{3N} \right).\label{res3}
\end{eqnarray}
This limiting result gives checkpoint  of our numerical calculations.

\subsection{Behavior under varying potential scattering}

Figure~\ref{res-tk} shows the temperature dependence of the normalized electric resistivity across the Kondo temperature.
For $|v|<v_{\rm cr} = 0.637$, the resistivity follows universal behavior as a function of $T/T_{\rm K}$.
As the potential scattering increases beyond the critical value,
the resistivity still follows the universal behavior 
in the Fermi-liquid range $T \ll T_{\rm K}$. 
As temperature increases, the resistivity starts to deviate from the universal curve around the Kondo temperature $T \sim T_{\rm K}$, and shows increasing behavior as the temperature is further increased. 
In the temperature range $T\gg T_{\rm K}$, we find that the resistivity for strong potential scattering can be described well with Kondo's formula\cite{kondo-1968} shown by dashed line in Fig.~\ref{res-tk}.
However, Kondo's formula cannot describe the properties for $T \le T_{\rm K}$.
Based on the numerical results for $T \ll T_{\rm K}$,
the coefficient $\alpha$ in eq.~(\ref{res2}) appears to be independent of $v$.

\begin{figure}
\centering
\includegraphics[totalheight=6cm,angle=0]{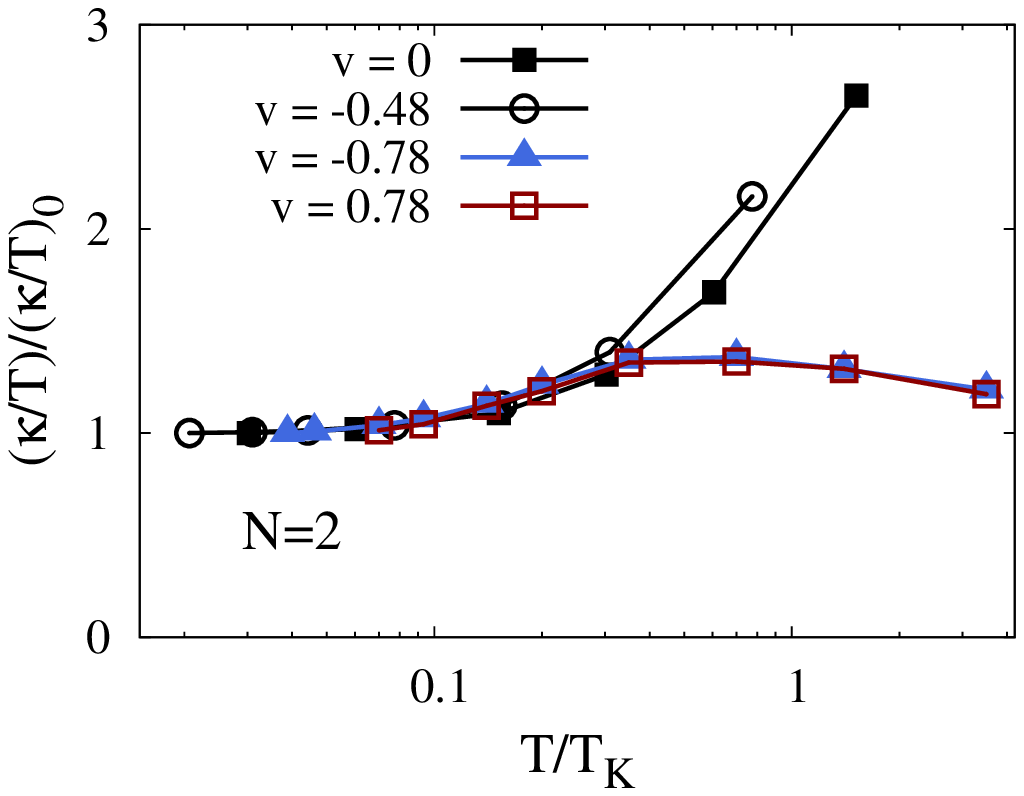}
\includegraphics[totalheight=6cm,angle=0]{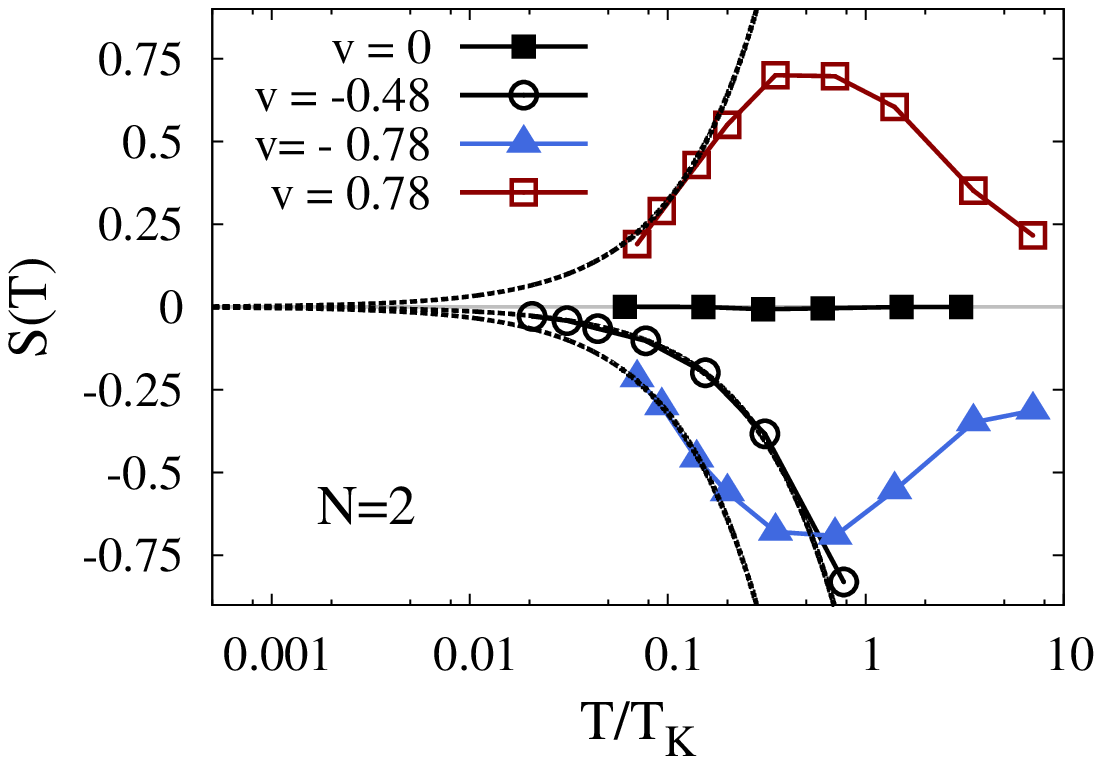}
\caption{
Temperature dependence of 
normalized thermal conductivity ({\sl upper}) and thermopower ({\sl lower})  
for the Kondo model with potential scattering for  $N=2$. 
Potential scattering $v$ and exchange $J$ are chosen as 
$v=0 \ (J=0.3)$, and $v=0.78, -0.48, -0.78 \ (J=0.44)$.
The {\sl dashed line} in the right panel corresponds to the Fermi-liquid behavior $S(T)\sim T$.
}\label{thermopower}
\end{figure}

The {\sl upper panel} of 
Fig.~\ref{thermopower} shows the temperature dependence of normalized 
thermal conductivity for $N=2$ under varying the potential scattering.
The thermal conductivity also follows the universal behavior even for large values of the potential scattering in the temperature range $T \ll T_{\rm K}$. 
Namely we obtain
\begin{align}
\frac{\kappa(T)}{T} / \left( \frac{\kappa(T)}{T} \right)_{0} = 1 + \gamma \left( \frac{T}{T_{\rm K}} \right)^2
\label{kappa2}
\end{align}
with $\gamma$ being a numerical constant independent of $v$.
Increasing further the temperature, thermal conductivity with large potential scattering highly deviates from the universal behavior. Namely, it decreases with increasing temperature for $T \gg T_{\rm K}$.

The {\sl lower panel} of 
Fig.~\ref{thermopower} shows the temperature dependence of 
thermopower for $N=2$ under varying the potential scattering.
The asymmetry of the impurity $t$-matrix is most reflected in the behavior of thermopower.
This is clear if we regard explicitly the expression for the integral $L_{1}$ given in eq.~(\ref{l-int}):
\begin{eqnarray}
L_{1} =  \frac{1}{2}\int_{-\infty}^{\infty} d \varepsilon \, 
 \frac{f^{\prime}(\varepsilon)\, \varepsilon}{{\rm Im}\,t(\varepsilon)}.
\label{l1-int}
\end{eqnarray}
Because of the factor $\varepsilon$ in the numerator, the thermopower measures the asymmetry in the energy dependence of ${\rm Im}\,t(\varepsilon)$.
Since the spectra is completely symmetric in the case of the ordinary Kondo model ($v=0$), the thermopower vanishes in this case as we obtain in the simulation (see Fig.~\ref{thermopower}).
On the other hand, the thermopower acquires strong temperature dependence when the potential scattering term is increased from $v=0$.
The thermopower in the simulation shows Fermi-liquid property 
\begin{align}
S(T) = \beta \left( \frac{T}{T_{\rm K}}\right) \label{S2}
\end{align}
for $T\ll T_{\rm K}$.
The coefficient $\beta$ is negative for $v<0$, while it is positive for $v>0$.
The different sign of $\beta$ can be simply understood if we recall that 
the asymmetry in the lineshape of ${\rm Im} \, t(\varepsilon)$ around the Fermi level is reversed 
against the sign change $v \rightarrow -v$.
Thus, the integrals given in eq.~(\ref{l-int}) have the properties 
$L_{0} (L_{2}) \rightarrow L_{0} (L_{2})$ and $L_{1}  \rightarrow - L_{1}$
under the sign change.
Namely, 
the electric resistivity $R(T)$ and thermal conductivity $\kappa(T)$ remain the same under $v \rightarrow -v$, while the thermopower changes sign as 
$S(T) \rightarrow - S(T)$.
We have indeed obtained these behaviors in the simulation as it can be seen in Figs.~\ref{res-tk} and \ref{thermopower}.

\subsection{Behavior under varying orbital degeneracy $N$}
\label{varying_N}

\begin{figure}
\centering
\includegraphics[totalheight=6cm,angle=0]{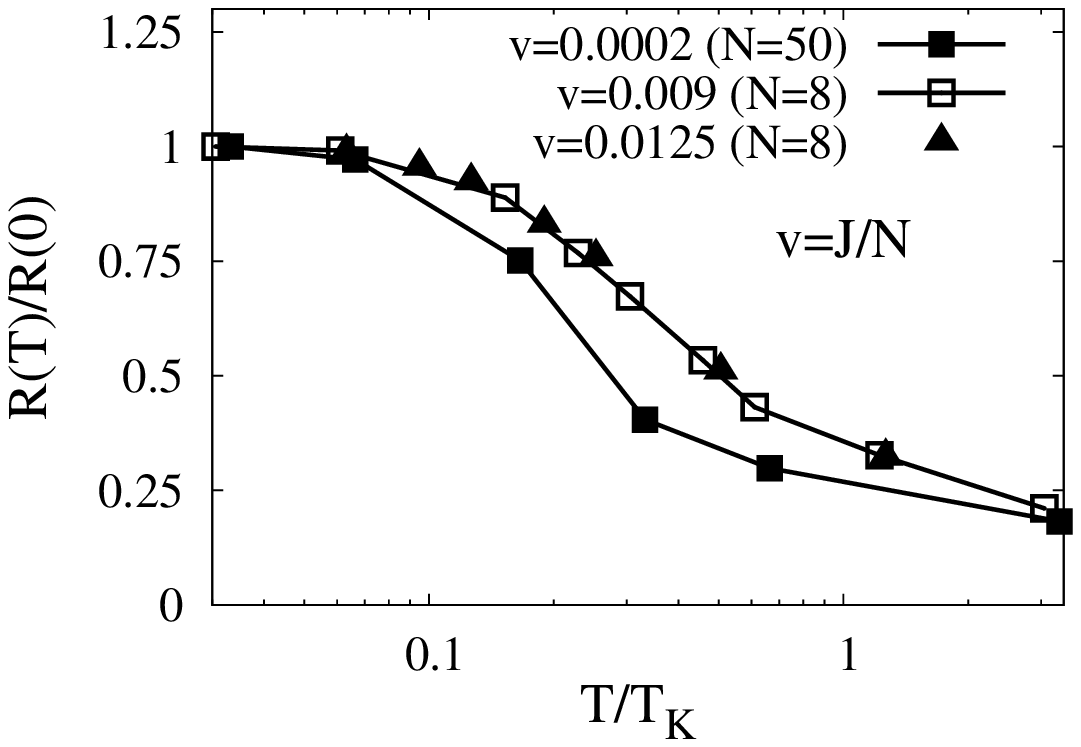}
\includegraphics[totalheight=6cm,angle=0]{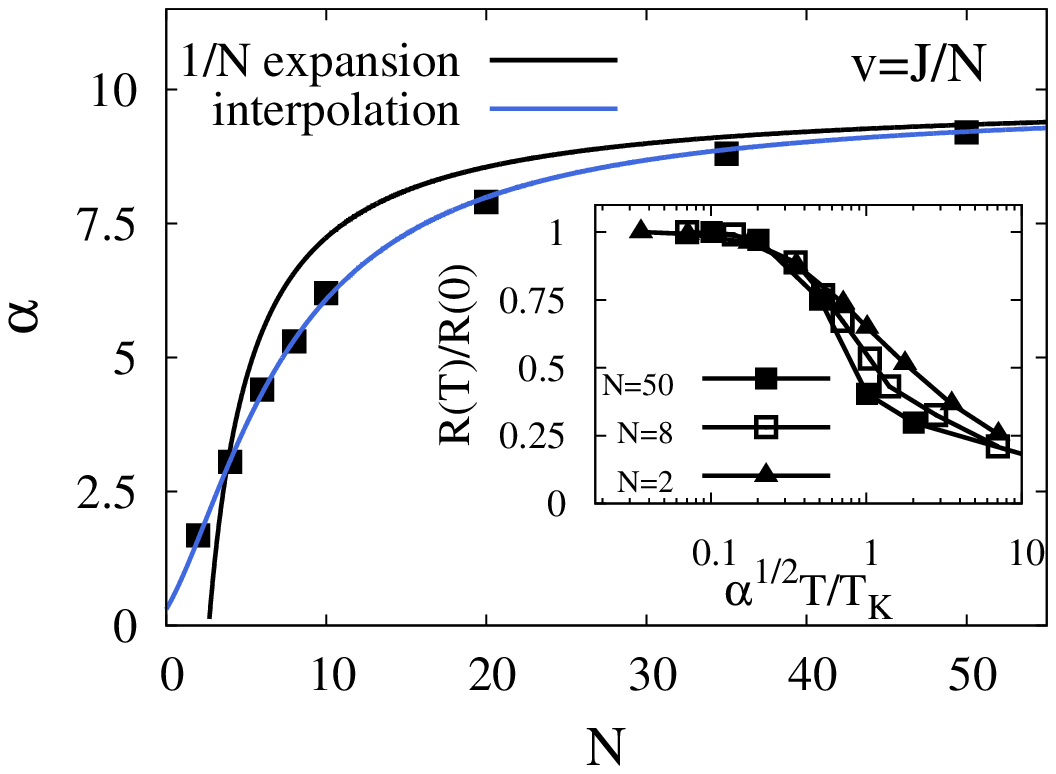}
\caption{\small {\sl Upper:} 
Temperature dependence of the normalized resistivity for the CS model (where $v=J/N$) for orbital degeneracies $N=8$ with $J=0.075$, $0.1$
 and $N=50$ with $J=0.0115$.
{\sl Lower:} Orbital degeneracy $N$-dependence of the coefficient $\alpha$ of the $T^2$ term in the low-temperature resistivity defined in eq.~(\ref{res2}).
The {\sl inset} shows the scaling behavior of the resistivity for different orbital degeneracies including $\alpha$.
}\label{resistivity-N}
\end{figure}

Let us now study the effect of orbital degeneracy.
The {\sl upper panel} of Fig.~\ref{resistivity-N} shows the temperature dependence of the normalized electric resistivity for large values of the orbital degeneracy for the CS model with $v=J/N$ across the Kondo temperature.
We observe again the universal behavior for a given value of the orbital degeneracy $N$.
In contrast to the behavior of the magnetic susceptibility shown in Fig.~\ref{susc}, 
the resistivity decreases monotonically as the temperature is increased even for large $N$.
In order to understand this feature, 
we assume that the $t$-matrix at low temperature
is determined by the quasi-particle density of states, which 
is approximately given by the effective Anderson model.
Namely we assume
\begin{align}
t(z) = V^2 G_f(z),
\end{align}
where $V$ is the effective hybridization and $G_f$ is the Green's function of the local electron in the effective Anderson model. Then
the Sommerfeld expansion of the conductivity leads to
\begin{eqnarray}
\sigma(T) &\sim & \int_{-\infty}^{\infty} \frac{1}{\rho_f(\varepsilon)}  \left( -\frac{\partial f(\varepsilon)}{\partial \varepsilon} \right) d\varepsilon \nonumber\\
 &=& 
\frac{1}{\rho_f( 0 )} \left( 1+ \frac{\pi^2}{6} (k_{\rm B}T)^2
 \left[ 2 \left( \frac{\rho_f^{\prime}(0)}{\rho_f(0)} \right)^2 - \frac{\rho_f^{\prime \prime}(0)}{\rho_f(0)}  \right] \right. \nonumber\\
&+& \left.  {\cal O}(T^4)  \right),\label{somm-sigma}
\end{eqnarray}
where
\begin{align}
\rho_f( \varepsilon)  = -\pi^{-1}{\rm Im}G_f(\varepsilon).
\end{align}
We obtain the resistivity $R(T)=\sigma(T)^{-1}$ from eq.~(\ref{somm-sigma}). 
Using the quasi-particle density of states for the non-interacting Anderson model with $v=0$,
we obtain
\begin{eqnarray}
R(T) \sim  \rho_f(0) \left[ 1- \frac{\pi^2}{3} (k_{\rm B}T)^2  \frac{1}{(
\eta ^2
 + \Delta^2)} +{\cal O}(T^4) \right],
\label{ressom}
\end{eqnarray}
where $\eta$ and $\Delta$ are the shift and the width of the resonance peak appearing in  $\rho_f(\varepsilon)$ at low temperatures.
Irrespective of the magnitude of parameters $\eta$ and $\Delta$, the coefficient of the $T^2$ term in the low-$T$ resistivity is always negative. Hence $R(T)$ given by eq.(\ref{ressom}) decreases as temperature increases for any value of $N$.
Namely, the quasi-particle picture is consistent with the monotonous change obtained in the simulation.

The resistivity obtained numerically has $T^2$ temperature dependence at low temperatures, which is expressed generally in eq.~(\ref{res2}).
The {\sl lower panel} of Fig.~\ref{resistivity-N} shows the coefficient $\alpha$ of the $T^{2}$ term for several values of the orbital degeneracy $N$ obtained in the simulation.
The result of $1/N$ expansion\cite{houghton-1987} given in eq.~(\ref{res3}) is also shown in the figure.
We find that the numerical data coincide with the $1/N$ expansion result for large values of the degeneracy $N$.
Thus, our CT-QMC simulation has produced accurate numerical results for large values of orbital degeneracies $N \rightarrow \infty$, which might be difficult in the case of other numerical techniques.
For small values of degeneracy $N$, the coefficient $\alpha$ shows linear-$N$ dependence in our simulation.

Now we fit the simulated $N$-dependence of 
$\alpha$ 
by a rational function as $\alpha (N)= F_{1}^{(l)}(N)/F_{2}^{(m)}(N)$,
where $F_{k}^{(l)} = c_{k0}+c_{k1}N+...c_{kl}N^{l}$.
We find that the minimal function which can give a good fit to the numerical data has the form
\begin{eqnarray}
\alpha (N)= \frac{c_{10}+c_{11}N+c_{12}N^2}{1+c_{21}N+c_{22}N^2}.\label{ratfit}
\end{eqnarray}
The limiting cases $N \rightarrow 0$ and $N \rightarrow \infty$ are obtained from the formula (\ref{ratfit}) as
\begin{eqnarray}
\alpha (N \rightarrow 0)\sim c_{10} + (c_{11}-c_{10}c_{21})N
\end{eqnarray}
and 
\begin{eqnarray}
\alpha (N \rightarrow \infty)\sim \frac{c_{12}}{c_{22}} \left[ 1 - \frac{1}{N} \left(\frac{c_{11}}{c_{12}}- 
\frac{c_{21}}{c_{22}}\right) \right].
\end{eqnarray}
Choosing the coefficients in eq.~(\ref{ratfit}) as
$c_{10}=0.3$; $c_{11}=0.53$; $c_{12}=0.17$; $c_{21}=0.1$; $c_{22}=0.0173$,
the numerical data can be fitted well (see Fig.~\ref{resistivity-N}).
The result of $1/N$ expansion\cite{houghton-1987} given in eq.~(\ref{res3})  is also reproduced in the large-$N$ range. 
In the inset of right part of Fig.~\ref{resistivity-N} we plot the normalized resistivity for different values of orbital degeneracy $N$ as a function of $\alpha (N)^{1/2}T/T_{\rm K}$.
The results show the scaling behavior of the resistivity in the Fermi-liquid range. For $T\ge T_{\rm K}$, the scaling property breaks down.

\begin{figure}
\centering
\includegraphics[totalheight=5.7cm,angle=0]{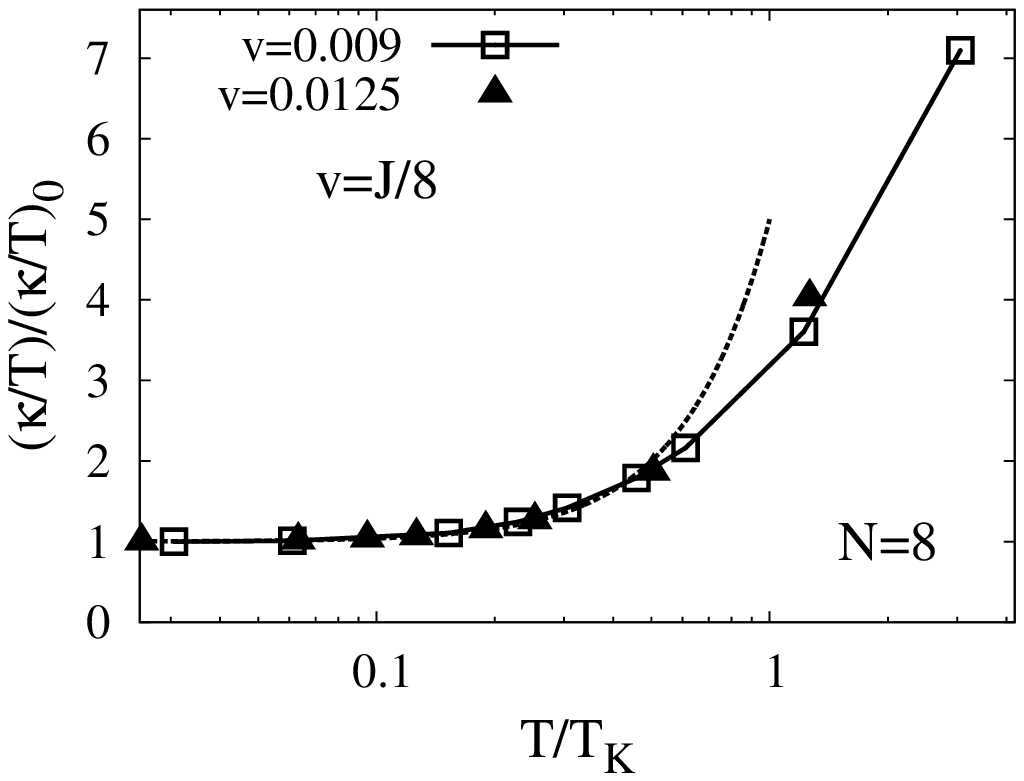}
\includegraphics[totalheight=5.7cm,angle=0]{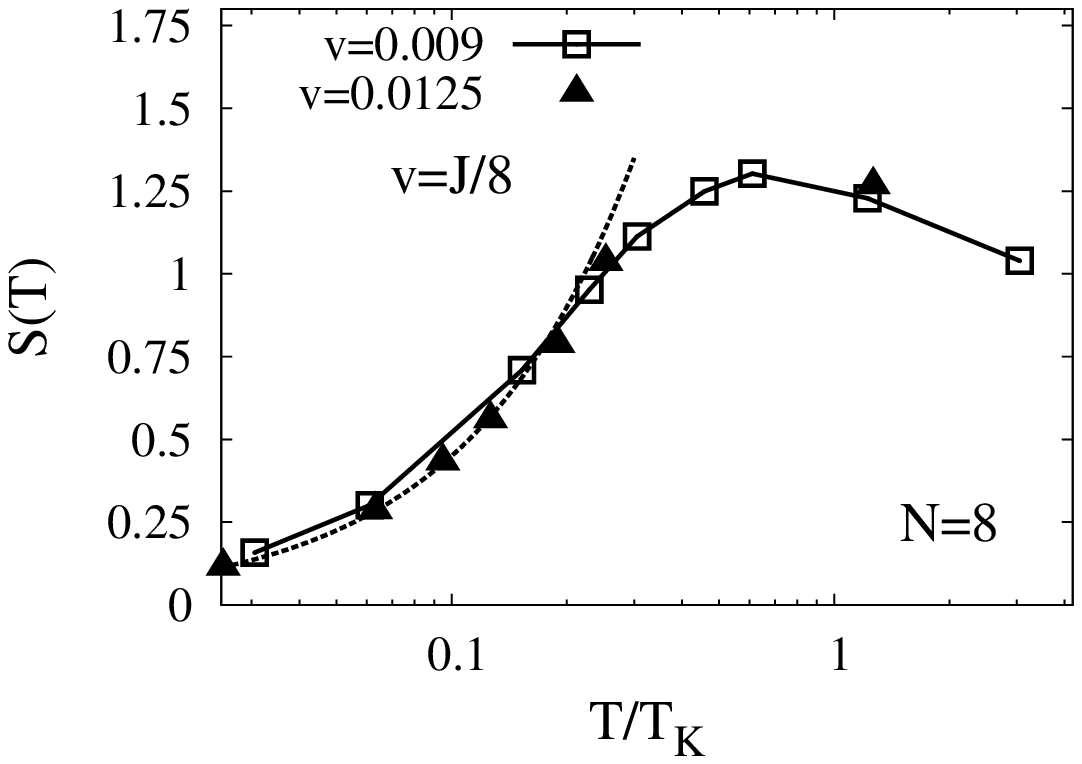}
\caption{\small 
Temperature dependence of 
normalized thermal conductivity ({\sl upper}) 
and thermopower ({\sl lower}) for the CS model (where $v=J/N$) for orbital degeneracy $N=8$. 
Potential scattering terms are chosen as $v=0.009$ ($J=0.075$) and $v=0.0125$ $J=0.1$). 
The {\sl dashed line} corresponds to the Fermi-liquid behavior given in eqs.~(\ref{S2}) and (\ref{kappa2}).}\label{transport-N8}
\end{figure}

In Fig.~\ref{transport-N8}, 
thermal conductivity $\kappa$ and
thermopower $S$  are shown for the CS model with orbital degeneracy $N=8$. 
The low-$T$ behaviors of $\kappa(T)$ and $S(T)$ are consistent with the Fermi-liquid result given in eqs.~(\ref{kappa2}) and (\ref{S2}).
As the temperature is further increased, the thermopower $S(T)$ has a peak, while the thermal conductivity $\kappa(T)$ monotonously
increases. 
Both quantities show universal behavior as a function of $T/T_{\rm K}$.

Note that 
the coefficient $\beta$ for $S(T)$ has a positive sign for large $N$.
This behavior is explained as follows.
In a manner similar to eq.~(\ref{somm-sigma}), $L_{1}$ is given by
\begin{eqnarray}
L_{1}(T) = \int_{-\infty}^{\infty} \frac{\varepsilon}{\rho_f(\varepsilon)} \left( - \frac{\partial f}{\partial \varepsilon} \right) = - \frac{2\pi^3}{3 \Delta} \eta T^2.
\end{eqnarray}
Using the result for $L_{0}$ given by eq.~(\ref{ressom}), we obtain $S(T)$  in the lowest order of $T$ as
\begin{eqnarray}
S(T) = \frac{2\pi^2}{3} \frac{\eta}{\eta^2+\Delta^2} T = \beta T.
\end{eqnarray}
Since 
$\eta\sim T_{\rm K}>0$ for large $N$, we obtain $\beta>0$. 
On the other hand, 
we obtain $\eta =0$ for the symmetric Anderson model with $N=2$.
In this case, the sign of $\beta$ depends on the sign of $v$
as it can be observed in Fig.~\ref{thermopower}.

\section{Quasi-particle properties}\label{section-fsr}
\subsection{Friedel sum rule}

At temperatures $T\ll T_{\rm K}$, the conduction electrons screen the magnetic impurity and they together form a local singlet. In this range the ground state is a local Fermi-liquid.
The Friedel sum rule (FSR) relates the phase shift for scattering of the conduction electrons 
by the impurity to its charge.
In the case of the CS model, the $f$-electron Green's function cannot be defined because there is no charge degrees of freedom in this localized model since it is eliminated.
Instead, the impurity $t$-matrix is used to describe the effect of exchange and potential scatterings. 
In Section \ref{section-transport}, we have related the $t$-matrix to the Green's function $G_{fv}(z)$ of localized electrons with potential scattering $v$.  It is the quantity $G_{fv}(z)$ that is expected to keep the FRS in the presence of $v$.

The FSR reads \cite{hewson-book}
\begin{eqnarray}
V^2G_{f}(0)  &=& - \frac{i}{\pi \rho_{0}} {\rm sin}^{2} \left( \frac{\pi}{N} \right), \label{fsr-rel-imre}
\end{eqnarray}
since the occupation number is unity in the CS and Kondo models.

\begin{figure}
\centering
\includegraphics[totalheight=4.7cm,angle=0]{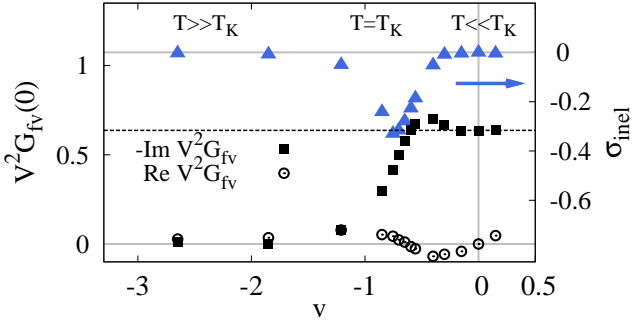}
\caption{Imaginary and real parts of the Green's function $G_{fv}$ expressed in eq.~(\ref{tz2}) at the Fermi level as a function of potential scattering $v$ together with the analytical result for $-{\rm Im}\, V^2 G_{fv}(0)$ obtained from the FSR ({\sl dashed line}).  
The figure also shows the inelastic scattering cross section $\sigma_{\rm inel}$ as blue triangles. The numerical data are obtained at temperature $\beta=1/T=1000$.}\label{gf-sineal}
\end{figure}

Figure~\ref{gf-sineal} shows the Green's function $G_{fv}(0)$ obtained by simulation at finite temperatures.
For small potential scattering the simulation results show good agreement with the expectation given in eq.~(\ref{fsr-rel-imre}).
As the value of $v$ is increased, ${\rm Re}\, V^2 G_{fv}(0)$ is still close to zero, but $-{\rm Im}\, V^2 G_{fv}(0)$ highly deviates from the theoretical result. 
The reason of this deviation is the following. 
The theoretical result given in eq.~(\ref{fsr-rel-imre}) is realized at $T=0$, 
which is almost realized for $T\ll T_{\rm K}$ in the simulation.
As the value of the potential scattering is increased, however, the corresponding Kondo temperature $T_{\rm K} = D {\rm e}^{1/(2\rho \widetilde{J})}$ rapidly decreases since the effective coupling $\widetilde{J}$ decreases rapidly\cite{kondo-1968}.
Therefore, we have to go at lower and lower temperatures in the simulation to achieve the condition $T\ll T_{\rm K}$, which is not fulfilled in Fig.~\ref{gf-sineal} for large values of $v$ since the results are obtained for a fixed temperature value $\beta = 1/T = 1000$.
In principle it is possible to go at lower temperatures in the simulation, but it becomes computationally harder.

\subsection{Optical theorem}

The optical theorem is less restrictive than the FSR since the former does not require the Fermi liquid ground state.
Optical theorem is related to the unitarity of the $S$ matrix,\cite{smatrix} and it follows when the scattering of the conduction electrons from the impurity is totally elastic at the Fermi level. 
Optical theorem was originally formulated for problems of scattering of a single-particle. When the scattering event happens without energy loss, i.e. it is totally elastic, there is a relation between the square and the imaginary part of the $t$-matrix. To express this relation, the $S$-matrix is decomposed as\cite{zarand-2004}
\begin{eqnarray}
S=1 + iT,\label{st}
\end{eqnarray}
where we write the matrix element of the $t$-matrix $T$ as $\langle n | T | n^{\prime} \rangle = 2\pi \delta(\varepsilon_{n} - \varepsilon_{n^{\prime}})\langle n | t | n^{\prime} \rangle$. Here, a state $| n \rangle$ represents a single-particle state with momentum ${\bf k}$ and spin $\sigma$ as $| n \rangle = | {\bf k} \sigma \rangle$.
After some manipulations we obtain from relation (\ref{st}) that
\begin{eqnarray}
\langle k | S S^{\dag} | k \rangle  - 1
&=& 2\pi \left[ 2\pi \sum_{n} \delta(\varepsilon_{k} - \varepsilon_{n} )  |\langle n | t | k \rangle|^{2} \right. \nonumber\\
 &+& \left. 2 {\rm Im} \,  \langle k | t | k \rangle \right].
\label{eqs4}
\end{eqnarray}
The first term of eq.~(\ref{eqs4}) in the right-hand side is related to the elastic scattering cross section $\sigma_{\rm el}$, while the second term to the total scattering cross section $\sigma_{\rm total}$ as\cite{zarand-2004, borda-2007}
\begin{eqnarray}
\sigma_{\rm el} &=& - 2 \pi \sum_{n} \delta(\varepsilon_{k} - \varepsilon_{n} )  |\langle n | t | k \rangle|^{2}
, \\
\sigma_{\rm total} &=& 2 {\rm Im} \,  \langle k | t | k \rangle.
\end{eqnarray}
The inelastic scattering cross section $\sigma_{\rm inel}$ is the difference of $\sigma_{\rm total}$  and $\sigma_{\rm el}$ as
\begin{eqnarray}
\sigma_{\rm inel} = \sigma_{\rm total}  - \sigma_{\rm el}.
\end{eqnarray}
For scattering only in the $s$ channel and assuming spin conservation, the inelastic cross section is expressed as\cite{borda-2007}
\begin{eqnarray}
\sigma_{\rm inel}(\omega) &=& (|s(\omega)|^2 - 1)/(2\pi) =  2 {\rm Im}\,  t(\omega) + 2 \pi \rho_{0} |t(\omega)|^2 \nonumber\\
&=& 
2 \pi \rho_{0} \left[\frac{1}{\pi \rho_{0}} {\rm Im}\,  t(\omega) + |t(\omega)|^2 \right],
\label{eqs5}
\end{eqnarray}
where $s$ and $t$ are eigenvalues of the $S$-matrix and $t$-matrix, respectively.

The eigenvalues $s$ of the $S$-matrix lie within the complex unit circle. 
The scattering is completely elastic, i.e. $\sigma_{\rm inel} = 0$, 
when the unitary condition $S^{\phantom{\dag}}S^{\dag}=1$ is satisfied, which means  $|s|^2=1$.
In this case 
we have the relation
\begin{eqnarray}
|t|^2 = - \frac{1}{\pi \rho_{0}} {\rm Im}\, t \label{eqs6}
\end{eqnarray}
from eq.~(\ref{eqs5}).

The problem is formulated so far for single-particle scattering, but it is more general and can be applied for many-particle problems as well such as the  single-channel Kondo model.
It is the most easily understood in the limit of $\varepsilon \rightarrow \infty$, when the magnetic impurity is completely decoupled from the conduction electrons. In this case the conduction electrons scatter without energy loss, and the optical theorem is held. Although the $t$-matrix is complicated and contains many scattering events at low temperatures, strictly in the limit of $\varepsilon \rightarrow 0$ the optical theorem holds again.
If we express the $t$-matrix as
\begin{eqnarray}
t = |t| {\rm e}^{i \theta},\label{gfshift}
\end{eqnarray}
where $\theta$ is the phase of the $t$-matrix $t$.
Equation (\ref{eqs6}) gives
\begin{eqnarray}
-{\rm Im}\, t(\varepsilon)  =  \frac{\sin^2 \theta}{\pi \rho_{0}}
 \label{imt0}
\end{eqnarray}
for the Kondo model with energy not only $\varepsilon=0$ but also $\varepsilon \rightarrow \infty$.
Namely, relation~(\ref{imt0}) is satisfied in the case when the scattering is totally elastic.
In the case of ordinary single-channel Kondo model ($v=0$), $\theta=-\pi/2$ at the Fermi energy, so from eq.~(\ref{imt0}) we recover the FSR
 \begin{eqnarray}
-{\rm Im}\, t(0)  =  \frac{1}{\pi \rho_{0}}.
 \label{imt0fsr}
\end{eqnarray}

Let us discuss the optical theorem in the context of our numerical data.
Figure~\ref{gf-sineal} shows the obtained inelastic scattering cross section $\sigma_{\rm inel}$ as a function of potential scattering.
Note that $\sigma_{\rm inel}$ shows non-monotonous behavior as a function of $v$.  Namely, $\sigma_{\rm inel}$ is almost
zero for small values of the potential scattering, but becomes non-zero as $|v|$ is increased.
With further increase of $|v|$, however, $\sigma_{\rm inel}$ approaches to zero again.  
Since $T_{\rm K}$ decreases as the potential scattering increases, 
in the large-$v$ range the condition 
$T\gg T_{\rm K}$ is satisfied 
at $T=10^{-3}$ used in the simulation.
On the other hand, 
in the small-$v$ range we have the condition
$T\ll T_{\rm K}$ with the same value: $T=10^{-3}$.
It is confirmed
 in the simulation that the optical theorem holds both at $T\ll T_{\rm K}$ and $T\gg T_{\rm K}$, but not for $T\sim T_{\rm K}$.
This happens because the ranges $T\ll T_{\rm K}$ and $T\gg T_{\rm K}$ correspond to the limits 
$\varepsilon/T_{\rm K} \ll 1$ and $\varepsilon /T_{\rm K} \gg 1$, respectively, where expression~(\ref{imt0}) is satisfied as explained above.

\section{Summary}\label{section-summary}

In this paper we have studied Kondo impurity models with potential scattering and orbital degeneracy by using CT-QMC numerical technique. We have derived 
accurate numerical results for
the impurity $t$-matrix, thermal, and transport properties in a wide temperature range across the Kondo temperature $T_{\rm K}$. 
Properties in the reverse Kondo range has been investigated in detail.
The results shown in this paper are numerically exact since CT-QMC does not use any approximation. 
We have explicitly demonstrated that CT-QMC simulation technique gives numerically exact results for large values of the orbital degeneracy $N$, which might be difficult to achieve in the case of other numerical techniques.

For large values of the potential scattering, 
non-trivial physics appears even in the impurity problem.
Namely, the resistivity shows anomalous increase with increasing temperature in contrast to the ordinary Kondo effect.
This unusual behavior is caused by an antiresonance developing around the Fermi energy in the quasi-particle density of states as the value of the potential scattering is increased.
This antiresonance does not influence the universal behavior of the magnetic susceptibility. 
However,
the sign of the Kondo logarithmic term changes in the resistivity when the potential scattering exceeds a critical value, i.e. in the reverse Kondo range, which causes the resistivity decrease with decreasing temperature.

We have studied the effect of strong potential scattering on thermal and transport properties of the Kondo impurity, and obtained that\\
(i) the magnetic susceptibility follows the universal temperature dependence even with strong potential scattering;\\
(ii) the resistivity also follows the universal temperature dependence for small values of the potential scattering;\\
(iii) when the potential scattering exceeds a critical value, the resistivity still shows universal behavior in the Fermi-liquid range, but starts to deviate from the universal curve around the Kondo temperature, and increases as the temperature is further increased;
(iv) the obtained temperature dependence of the resistivity for $T \gg T_{\rm K}$ in the reverse Kondo range agrees quantitatively with Kondo's theoretical result;\\
(v) the thermal conductivity also shows universal behavior in the Fermi-liquid range, but highly deviates from the universal curve for $T \gg T_{\rm K}$ in the reverse Kondo range; \\
(vi) the asymmetry of the $t$-matrix developing with increasing value of the potential scattering is most reflected in the temperature dependence of the thermopower;\\
(vii) the sign of the thermopower depends on the sign of the potential scattering.

In addition to the study of thermal and transport properties, we have discussed the Friedel sum rule and optical theorem as well.
We have shown that the $t$-matrix of the Kondo model in the presence of potential scattering is not the relevant quantity for the Friedel sum rule.
Instead, the Friedel sum rule is satisfied with a proper limit of the $f$-electron Green's function.
We have demonstrated that optical theorem is less restrictive than the Friedel sum rule, because 
the former
holds not only in the Fermi-liquid range, but for large energies as well.

Finally we mention 
an interesting question
 whether the behavior found for strong potential scattering has relevance in real systems.
Note that recent STM experiments on URu$_2$Si$_2$
have found that the density of states shows Fano lineshape in the normal phase, and in the dilute system U$_{x}$Th$_{1-x}$Ru$_{2}$Si$_{2}$ the resistivity decreases with decreasing temperature.
Since some important aspect of the U ion with non-Kramers configuration $5f^2$ may not be described by a localized spin with $S=1/2$,
account of the strong potential scattering in more realistic models is desirable.
We hope that our results in this paper will stimulate further study
concerning the Fano lineshape and other aspects, which reflect interplay of the Kondo effect and potential scattering.

\acknowledgements

The authors are grateful to Dr. J. Otsuki for his guidance on the details of CT-QMC simulation technique, and also for useful discussions.
AK acknowledges the Magyary programme and the EGT Norway Grants, and also the financial support from the European Union Seventh Framework Programme through the Marie Curie Grant PIRG-GA-2010-276834.


\begin{thebibliography}{99}

\bibitem{coles64}
B.R. Coles, Phys. Letters {\bf 8}, 243 (1967).

\bibitem{lieke78}
W. Lieke, J.H. Moeser and F. Steglich, 
Z. Physik  B {\bf 30}, 155 (1978).

\bibitem{schmid80} 
W. Schmid, E. Umlauf, F. Steglich and P. Thalmeier, 
Solid State Commun. {\bf 35}, 325 (1980).

\bibitem{lieke80}
W. Lieke, F. Steglich, K. Rander and H. Keiter, 
Phys. Rev. B {\bf 20}, 2129 (1979).

\bibitem{fischer67}
K. Fischer, Phys. Rev. {\bf 158}, 613 (1967).

\bibitem{kondo-1968}
J. Kondo, Phys. Rev. {\bf 169}, 437 (1968).

\bibitem{schmidt-2010}
A. R. Schmidt, M. H. Hamidian, P. Wahl, F. Meier, A. V. Balatsky, J. D. Garrett, T. J. Williams, G. M. Luke, and J. C. Davis, Nature {\bf 465}, 570 (2010).

\bibitem{amitsuka-1994}
H. Amitsuka and T. Sakakibara, J. Phys. Soc. Jpn. {\bf 63}, 736 (1994).

\bibitem{otsuki-2007}
J. Otsuki, H. Kusunose, P. Werner, and Y. Kuramoto, J. Phys. Soc. Jpn. {\bf 76}, 114707 (2007).

\bibitem{otsuki-2009a}
J. Otsuki, H. Kusunose, and Y. Kuramoto, J. Phys. Soc. Jpn. {\bf 78}, 014702 (2009).

\bibitem{otsuki-2009b}
J. Otsuki, H. Kusunose, and Y. Kuramoto, J. Phys. Soc. Jpn. {\bf 78}, 034719 (2009).

\bibitem{coqblin-schrieffer-1969}
B. Coqblin and J. R. Schrieffer, Phys. Rev. {\bf 185}, 847 (1969).

\bibitem{hewson-book}
A. C. Hewson: The Kondo problem to Heavy Fermions (Cambridge University Press, Cambridge, 1993).

\bibitem{fano-1961}
U. Fano, Phys. Rev. {\bf 124}, 1866 (1961).

\bibitem{mahan-book}
G. D. Mahan: Many-Particle Physics, 3rd edition (Plenum, New York, 2000).

\bibitem{pruschke-2006}
C. Grenzebach, F. B. Anders, G. Czycholl, and T. Pruschke, Phys. Rev. B {\bf 74}, 195119 (2006).

\bibitem{hamann-1967}
D. R. Hamann, Phys. Rev. {\bf 158}, 570 (1967).

\bibitem{Nozieres-1974}
P. Nozi$\grave{\rm e}$res, J. Low Temp. Phys. {\bf 17}, 31 (1974).

\bibitem{houghton-1987}
A. Houghton, N. Read, and H. Won, Phys. Rev. B {\bf 35}, 5123 (1987).

\bibitem{smatrix}
The $S$-matrix is given by 
$S=T_{\tau} {\rm exp}[\int_{-\infty}^{\infty} {\rm H}_{\rm int} (\tau)d\tau ]$ in the interaction representation, where $T_{\tau}$ is the time-ordering operator with $\tau$ being the imaginary time.

\bibitem{zarand-2004}
G. Zar\'and, L. Borda, J. Delft, and N. Andrei, Phys. Rev. Lett. {\bf 93}, 107204 (2004).

\bibitem{borda-2007}
L. Borda, L. Fritz, N. Andrei, and G. Zar\'and, Phys. Rev. B {\bf 75}, 235112 (2007).



\end{thebibliography}
\end{document}